\documentclass{article}
\usepackage{amsmath,amssymb}
\usepackage{stmaryrd}
\usepackage{enumerate}
\usepackage{array}
\usepackage{bbm}
\usepackage{eqparbox}
\usepackage{frame}
\usepackage{float}
\usepackage{algorithm}
\usepackage{color}
\usepackage{url}
\usepackage{epsfig}
\usepackage{multirow}
\usepackage{color}
\usepackage{subfig}
\usepackage{cite}
\usepackage{todonotes}
\usepackage[font=small]{caption}
\usepackage{balance}
\usepackage{graphicx}
\usepackage{adjustbox}
\usepackage{geometry}
\usepackage{spconf}

\newcommand{\beq}{\begin{equation}}
\newcommand{\eeq}{\end{equation}}
\newcommand{\bdm}{\begin{displaymath}}
\newcommand{\edm}{\end{displaymath}}

\newtheorem{definition}{Definition}

\newcommand{\bd}{\begin{definition}}
\newcommand{\ed}{\end{definition}}

\newcommand{\bv}{\begin{vugraph}}
\newcommand{\ev}{\end{vugraph}}
\newcommand{\bi}{\begin{itemize}}
\newcommand{\ei}{\end{itemize}}
\newcommand{\ben}{\begin{enumerate}}
\newcommand{\een}{\end{enumerate}}

\newcommand{\bean}{\begin{eqnarray*} }
\newcommand{\eean}{\end{eqnarray*} }
\newcommand{\bea}{\begin{eqnarray} }
\newcommand{\eea}{\end{eqnarray} }

\newcommand{\ba}{\begin{array} }
\newcommand{\ea}{\end{array} }

\newcommand{\Section}[1]{ \section{#1}} 

\title{Hieroglyph: Hierarchical Glia Graph Skeletonization and Matching}

\name{Tiffany T. Ly$^{\dagger}$, Tamal Batabyal$^{\dagger}$, Jeremy Thompson$^{\ddagger}$, Tajie Harris$^{\ddagger}$, Daniel S. Weller$^{\dagger}$, and Scott T. Acton$^{\dagger}$}
\address{$^{\dagger}$C.L. Brown Department of Electrical \& Computer Engineering \\ $^\ddagger$ Center for Brain Immunology and Glia, Department of Neuroscience, University of Virginia\\
Charlottesville, Virginia USA}

\begin{document}
\pagenumbering{gobble}
\maketitle

\begin{abstract}
Automatic 3D reconstruction of glia morphology is a powerful tool necessary for investigating the role of microglia in neurological disorders in the central nervous system. Current glia skeleton reconstruction techniques fail to capture an accurate tracing of the processes over time, useful for the study of the microglia motility and morphology in the brain during healthy and diseased states. We propose \textit{Hieroglyph}, a fully automatic temporal 3D skeleton reconstruction algorithm for glia imaged via 3D multiphoton microscopy. \textit{Hieroglyph} yielded a 21\% performance increase compared to state of the art automatic skeleton reconstruction methods and outperforms the state of the art in different measures of consistency on datasets of 3D images of microglia. The results from this method provide a 3D graph and digital reconstruction of glia useful for a myriad of morphological analyses that could impact studies in brain immunology and disease. 
\end{abstract}


\Section{Introduction}
Microglia are the tissue resident immune cells of the brain parenchyma and play an active role in brain homeostasis.  The advancements made in the recent studies of microglia have shifted our understanding of the impact of microglia not only in development, but also its role in injury, diseases, and aging \cite{colonna2017microglia, schafer2015microglia, poliani2015trem2, cronk2018peripherally}. Microglia processes (thin 'legs' that extend from the cell body) are constantly in motion for surveillance under homeostasis, which is predicted to allow for microglia to sense and respond rapidly to their environment \cite{wake2009resting, tremblay2010microglial, davalos2012fibrinogen,madry2017microglial}. During brain injury and disease this continual movement is altered as microglia retract their processes and take on a more amoeboid morphology.  However, little is known about how the decrease in microglia processes activity and their motility affect their ability to perform surveillance functions in the brain. 

The complexity of glia morphology makes it difficult to automate the analysis of glia motility. Existing studies have manually traced glial images or used heuristic image processing methods to measure process length, extension and retraction over time \cite{nimmerjahn2005resting, wu2008resting, davalos2005atp}. Researchers have developed automatic image analysis methods involving the reconstruction of skeletons of the microglia processes \cite{young2018quantifying, paris2018promoij}. In \cite{young2018quantifying}, the skeletonization was semi-automatic in that the user went through many preprocessing tasks in ImageJ before achieving a 2D skeleton. However, 2D skeletonization loses information since the skeletons may overlap in the z direction, as shown in our experimental results. ProMoIJ achieves an automatic reconstruction of a 3D skeleton of glia, which is then used to analyze microglia motility\cite{paris2018promoij}. However, the skeletonization is not accomplished for the entire cell, rather each process of the glial cell is manually selected by the user. Furthermore, the user must define a set of parameters to heuristically preprocess the image and create a skeleton. This reconstruction of the processes over time are manually assisted. In this paper, we describe \textit{Hieroglyph} -- a hierarchical glia graph skeletonization and matching system. 

\begin{figure}[t!]
	\centering
  \includegraphics[width=.9\linewidth]{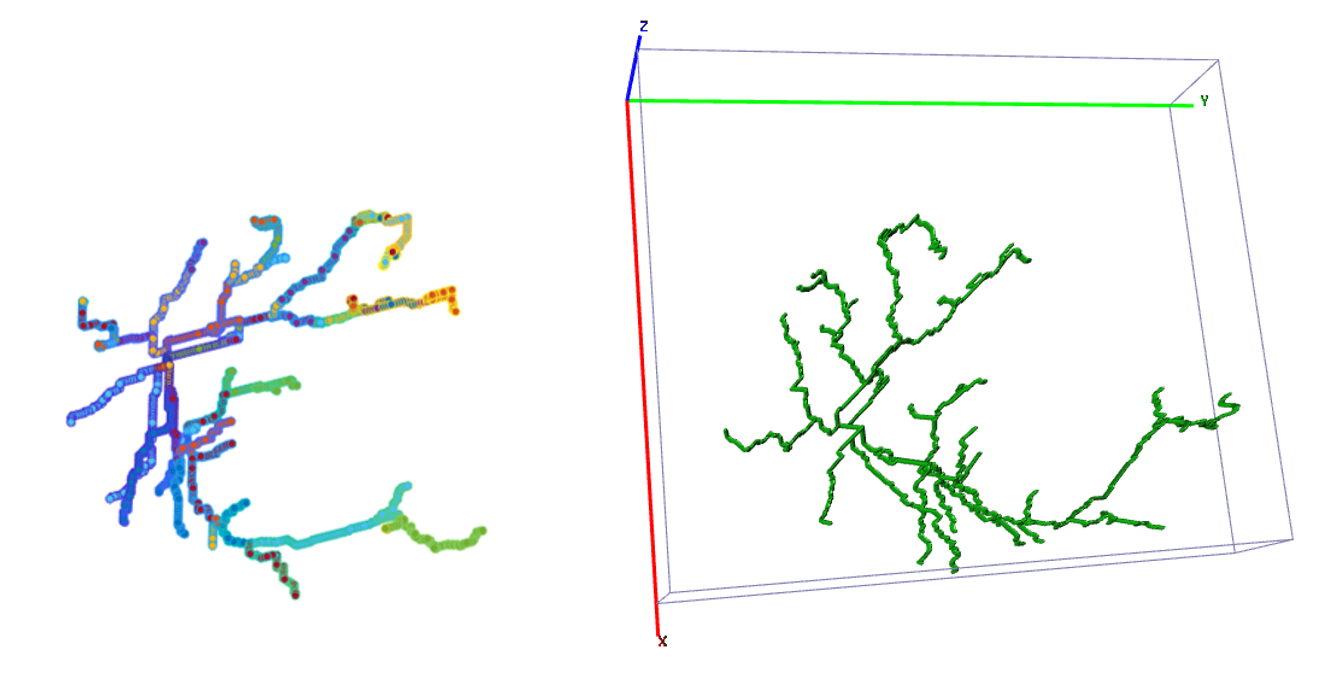}
 
  \caption{\small{Hierarchy of a glial cell shown via color from violet to red (left) and 3D skeleton of the cell (right). }}
   \label{fig:hierd}
\vspace{-0.5cm}  	
\end{figure} 

Morphological reconstruction is also an important technique for the analysis of neuron morphology. Yet even in NeuroMorpho.org, the largest curated inventory of publicly accessible 3D neuronal reconstructions, less than 5\% of the reconstructions are traced in a semi-automatic fashion, while the remainder are manually traced \cite{donohue2011automated}. State-of-the-art methods for automatic skeletonization include medial axis thinning, which involves iteratively eroding the boundary of the object until only a one-voxel-thick limb remains. The difficulty in achieving an accurate skeleton still remains, as the existing methods are dependent on the object shape or the image intensity variations. Non-smooth and irregular structure can lead to spurious edges, false branches, discontinuities, loops and other anatomical or structural inaccuracies in the skeleton reconstruction. Such methods typically require a prior segmentation of the object of interest, which makes the resulting skeleton extremely dependent on the segmentation. The inaccuracies of the prior information can lead to inconsistent and erroneous skeletonization.

In this paper, we propose, \textit{Hieroglyph}, a hierarchical algorithm for an automatic 3D reconstruction of consistent skeletons over time using glial images. First, we provide an automatic tracing method of a glial cell that achieves an accurate 3D skeleton without loops or discontinuities. Second, we use the graph representation of the prior skeleton information to achieve consistent 3D skeletons of a microglia over time. Third, our 3D temporal reconstructions are represented in digital and graph format that can be easily manipulated by statistical and graphical morphology analysis.  

\Section{Methods}
\textit{Hieroglyph} produces a consistent temporal digital reconstruction of a glia skeleton by using prior information from previous 3D acquisition. The skeleton from a previous acquisition is evolved by representing the cell as a graph where the glia process lengths are stored as edge information and each bifurcation is stored as a node. In Section \ref{sec:dijkstra}, we describe the use of the graph representation of the glia to achieve a skeleton tracing using a shortest path algorithm. In Section \ref{sec:temporal}, \textit{Hieroglyph} evolves the previous skeleton to match the image in the next time frame. Each generated skeleton is employed to create another consistent 3D skeleton for the following glia image in the time series. 

\begin{figure}[h]
	\centering
  \includegraphics[width=\linewidth]{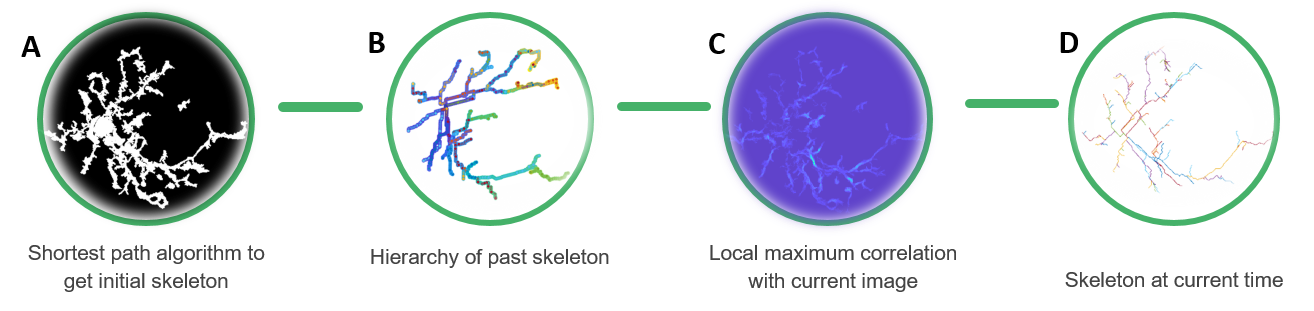}
  \caption{\small{Overall methodology. A) The shortest path is take from the end node to the soma to get the first skeleton. B-C) a skeleton is morphed to another in an hierarchical manner D) the resulting skeleton. (B-C) is repeated for the remaining images in the time series. }}
   \label{fig:method}
\vspace{-0.7cm}  	
\end{figure} 

\subsection{3D skeleton: shortest path in a graph} \label{sec:dijkstra}
Let us consider a set of 3D time series images where the segmentation of the image at time \textit{t}=1 is represented as an adjacency matrix of an undirected, weighted, rooted tree graph, \textit{Adj(G)}. The graph consists of vertices and edges, \textit{G=(V,E)}, where the vertices are initialized at every foreground pixel in the segmented image and size of \textit{V} equals the number of foreground voxel in the segmented image. The edges are weighted by the spatial Euclidean distance between the voxels, $e = 1, \sqrt{2}, or \sqrt{3}$  \cite{mukherjee2013tree2tree2, peng2010automatic}. The adjacency matrix is filled with the weights of the edges between all the foreground voxels. The size of \textit{Adj(G)} is $N \times N$ where $N$ is the number foreground pixels, or the number of vertices. 

From a biological standpoint, we know that our reconstructed graph is a simple graph which should not contain any loops or discontinuities from the processes to the soma of the glia. Thus, to construct our tracing of the cell, we use Dijkstra's algorithm \cite{dijkstra1959note} to find the shortest path between the terminal nodes of the processes to the soma of the glia. The terminal nodes are extracted from the segmentation of the 3D glia, and the soma vertex is the center of mass of the 3D soma segment. The algorithm starts at terminal node and finds the shortest route within the given adjacency matrix of the graph to the soma, or root node. The route of the voxels between the soma and  the terminal nodes result in a 3D skeleton tracing of the glia. \textit{The benefit of the graph representation is the rich information provided that includes the hierarchy of the processes, the bifurcation points, and the endpoints. }These properties are exploited in the creation a consistent skeleton for the subsequent glia image in a time series.
\begin{figure*}[t!]
	\centering
	\renewcommand{\tabcolsep}{0.05cm}
	\setlength{\belowcaptionskip}{-10pt}
	\begin{adjustbox}{width=\textwidth}
	{
			\begin{tabular}{ccccc}
			
				\multicolumn{5}{c}{Ground truth}
                \\
				\includegraphics[width=.35\linewidth, height = 0.25\linewidth, scale=0.15]{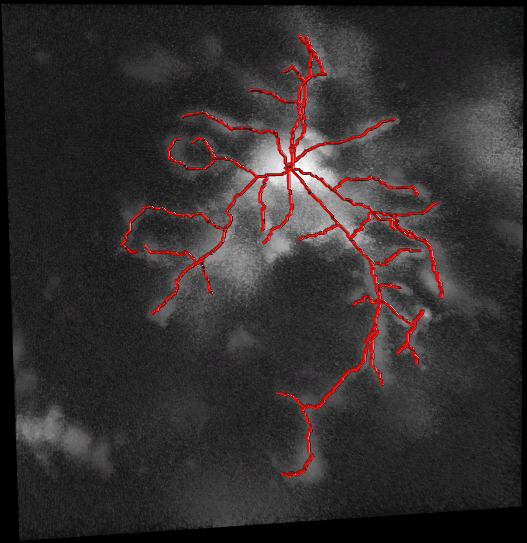} &
				\includegraphics[width=.35\linewidth, height = 0.25\linewidth, scale=0.15]{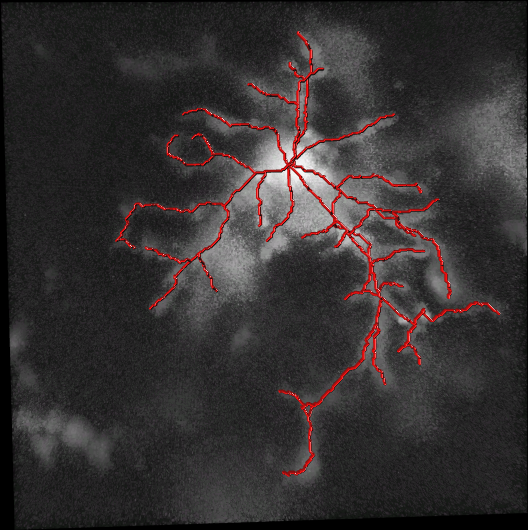} &
				\includegraphics[width=.35\linewidth, height = 0.25\linewidth, scale=0.15]{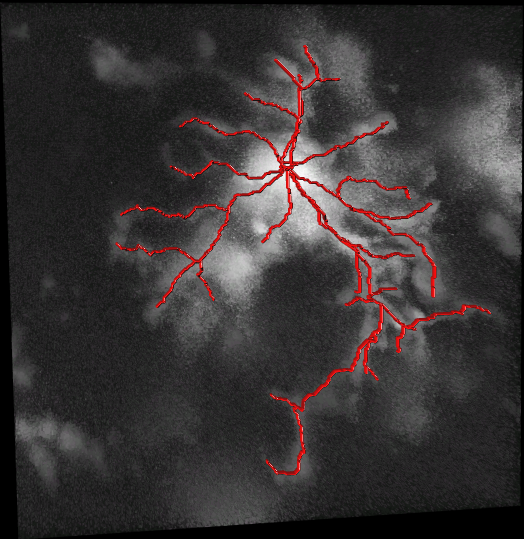} &
                \includegraphics[width=.35\linewidth, height = 0.25\linewidth, scale=0.15]{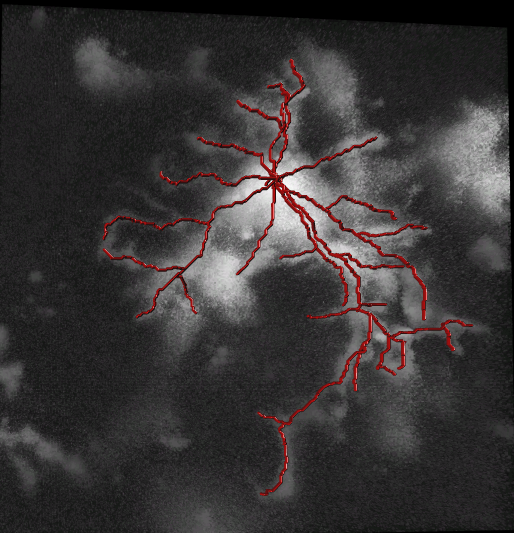} &
				\includegraphics[width=.35\linewidth, height = 0.25\linewidth, scale=0.15]{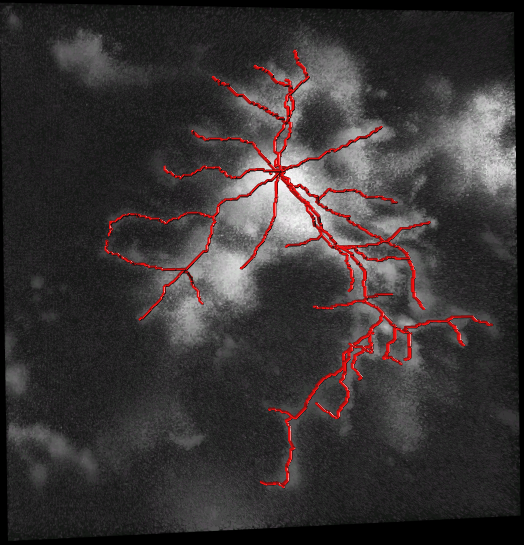} 
				\\
				\multicolumn{5}{c}{Hieroglyph}
                \\
				\includegraphics[width=.35\linewidth, height = 0.25\linewidth, scale=0.1]{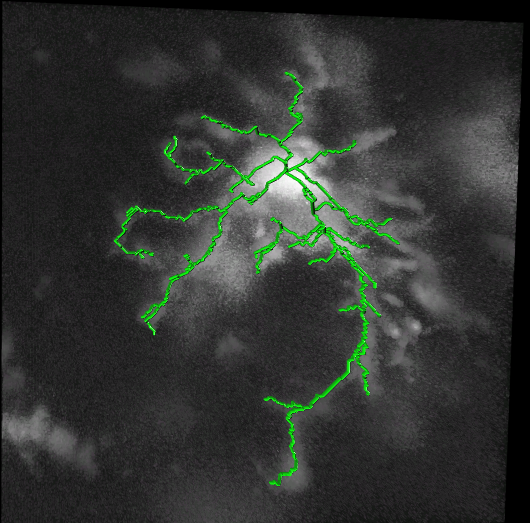} &
				\includegraphics[width=.35\linewidth, height = 0.25\linewidth, scale=0.1]{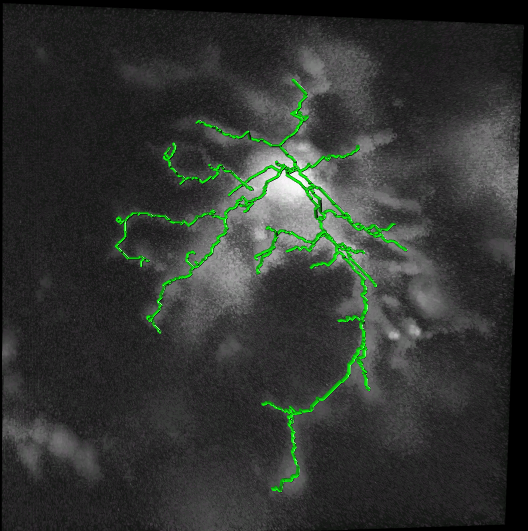} &
				\includegraphics[width=.35\linewidth, height = 0.25\linewidth, scale=0.1]{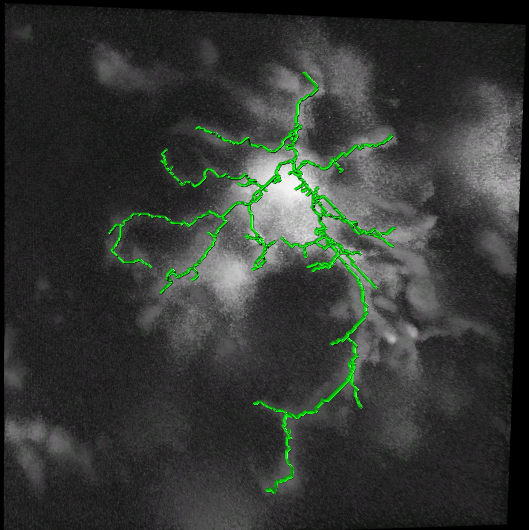} &
                \includegraphics[width=.35\linewidth, height = 0.25\linewidth, scale=0.1]{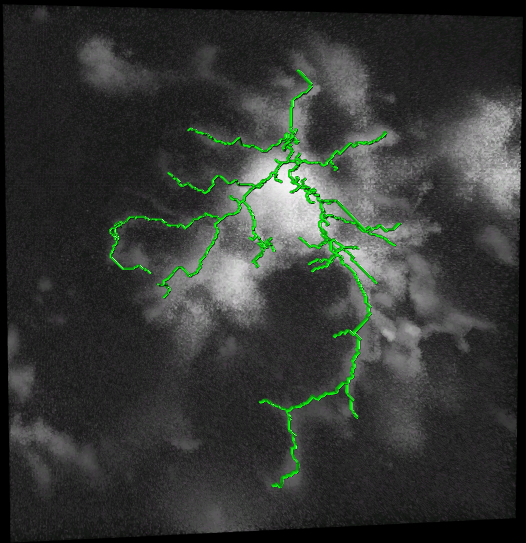} &
				\includegraphics[width=.35\linewidth, height = 0.25\linewidth, scale=0.1]{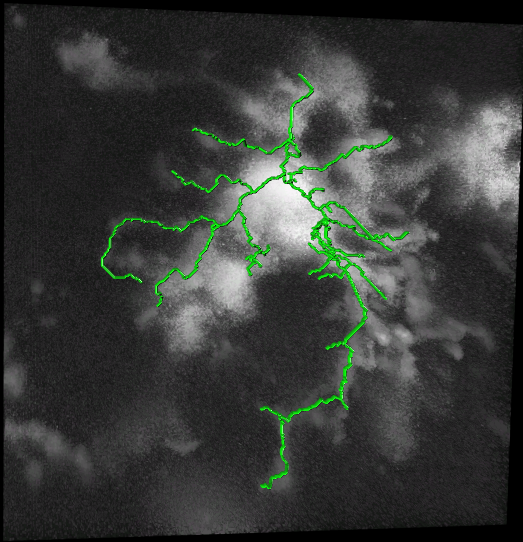} 
				\\
				\multicolumn{5}{c}{Skel2Graph\cite{kollmannsberger2017small}}
                \\
			\includegraphics[width=.35\linewidth, height = 0.25\linewidth, scale=0.1]{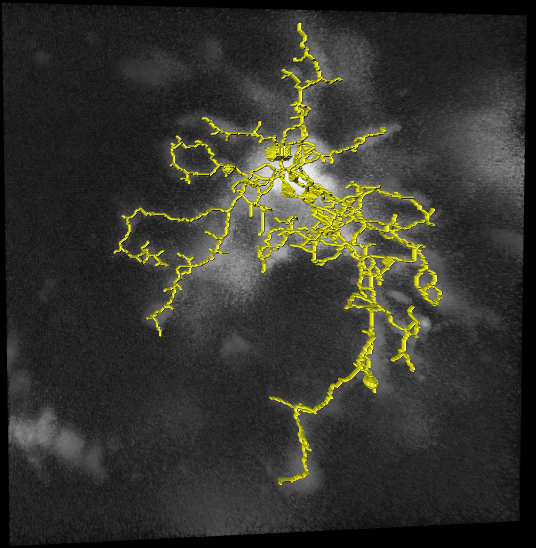} &
				\includegraphics[width=.35\linewidth, height = 0.25\linewidth, scale=0.1]{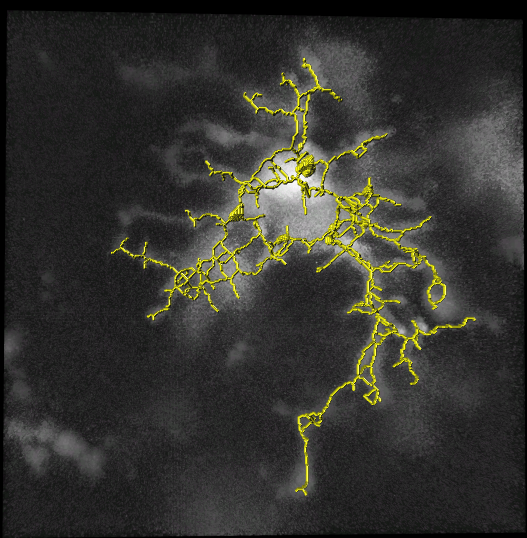} &
				\includegraphics[width=.35\linewidth, height = 0.25\linewidth, scale=0.1]{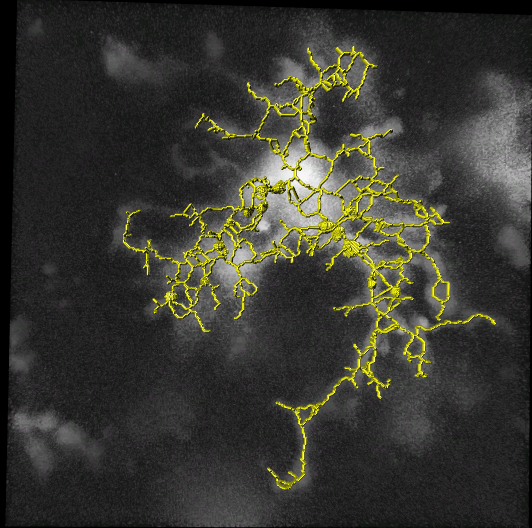} &
                \includegraphics[width=.35\linewidth, height = 0.25\linewidth, scale=0.1]{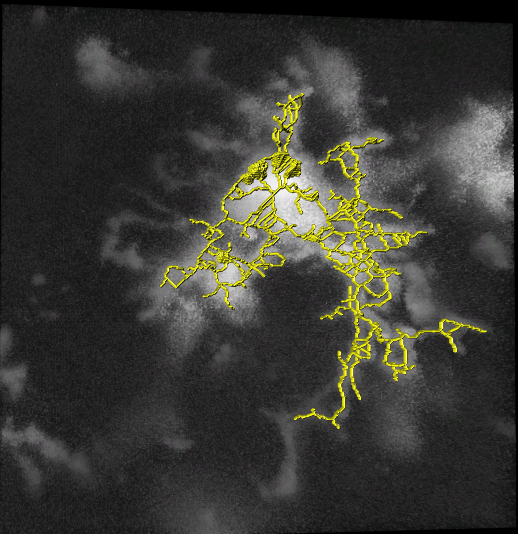} &
				\includegraphics[width=.35\linewidth, height = 0.25\linewidth, scale=0.1]{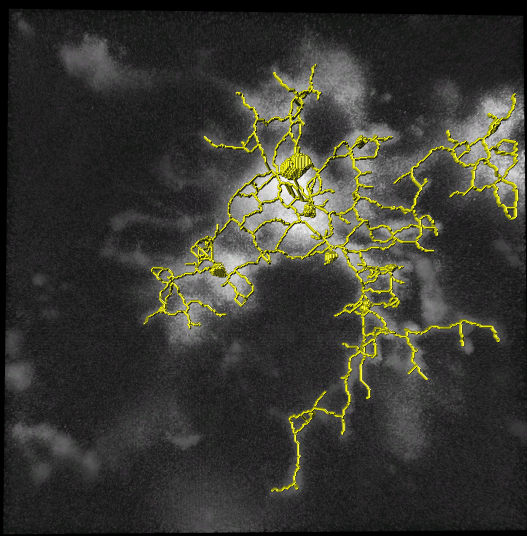} 
				\\
                 {time \textit{t}=1 minute} & {time \textit{t}=2 minute} & {time\textit{ t}=6 minute} &{time\textit{ t}=10 minute} & {time\textit{ t}=13 minute}
				
			\end{tabular}
		}	
	\end{adjustbox}
	\caption{\small{Segmentation results of 3D microglia images.}}
	
	\vspace{-0.5cm}
	\label{fig:results}
\end{figure*}

\subsection{Consistent 3D skeletons from temporal information}\label{sec:temporal}
Acquiring glial skeletons solely from segmentation can result in inconsistencies between acquisitions in time and is computationally burdensome. A single glia cell over time extends and retracts the processes while keeping the same number of branches that emerge from the soma. Thus the morphology of a glial cell between subsequent acquisitions is consistent. We propose a method that uses prior temporal information combined with intensity information from the current image. \textit{Hieroglyph} seeks to drive the skeleton from a previous time frame to the vessel-like information in the original image of the current time stack. 

The latter information is gathered by using the Hessian-based vessel enhancement technique to distinguish tubular structures in an image. This technique utilizes a multiscale function according to three direction of the orthonormal eigenvectors, $\textbf{e}_i(\textbf{x})$, where \textbf{x} is the pixel position within the image domain \cite{frangi1998multiscale,sato1998three}. These directional eigenvectors are attained by computing the Hessian matrix of a Gaussian smoothed 3D image and then ordering the eigenvectors by the increasing magnitudes of the eigenvalues $|\lambda_1| \leq |\lambda_2| \leq |\lambda_3| >>0$. A vessel-enhanced image is obtained with a low $|\lambda_1|$ value and high $|\lambda_2|$ and $|\lambda_3|$ magnitudes. We call the vessel enhanced image $I_v$. 

The initial skeleton from the previous time frame, $S_{t-1}$ is broken into hierarchies, where the root node is equal to 0 and the hierarchy increases towards the terminal branches. Every bifurcation of a process separates the process into another hierarchy $H$, where $H={h_1,h_2,...h_i}$ is a set of hierarchies, length $i$. Every segment belongs to a cluster $h_i$ within the set. The algorithm morphs the skeleton, one segment at a time, starting from the lowest hierarchy until it achieves the maximum response with the vessel enhanced image. This is repeated for all segments of the skeleton in $H$. The final skeleton at time \textit{t} is given by
\begin{equation}
S_t = \max  \sum_{h=1}^i S_{t=1}(h^*) \times I_v
\end{equation}
\begin{equation}
I_v = 
 \begin{cases} 
      x & x> 0, \\
     -x_{avg} & x\leq 0 .
 \end{cases}
\end{equation}
where $h^*$ is the morphed segment from the previous skeleton and $x$ is the voxel value in $I_v$. The morphing of the segments are changes in the 26 cardinal directions. The morphed segments are bounded by the following conditions: 1) The first hierarchy must start at the root node. 2) Segments may not overlap with each other (no loops). 3) The bifurcation points are regularized so they do not drastically move. 

The zero intensity values in $I_v$ are set to the negative value of the average pixel intensity to penalize morphing beyond vessels in the Hessian map. Once a new segment is created, the tree is rerouted resulting in an updated graph representation of the skeleton so that the routes and bifurcation nodes are updated. 

\Section{Experimental Results and Analysis}
\subsection{Imaging and fluorescence technique}
The dataset consists of 3D images of microglia from living mice using multiphoton microscopy. To label microglia in the mouse brain, we used mice with an inducible Cre recombinase under the control of the CX3CR1 promoter crossed to the Ai6 fluorescent reporter mouse  (Jackson Laboratories, Bar Harbor, ME) to generate CX3CR1creERT2/+ X Ai6ZsGreen \cite{yona2013fate, madisen2010robust}. At post-natal day (P23) 23, mice were given 10uL/g body weight of a 20mg/mL Tamoxifen (Sigma) solution in corn oil to induce recombination of the floxed stop codon leading to ZsGreen expression in microglia. All procedures adhered to guidelines of the Institutional Animal Care and Use Committee (ACUC) at the University of Virginia. Microglia of adult mice (7-10 weeks old) were imaged using a Leica TCS SP8 multiphoton microscopy system equipped with a Coherent Chameleon Ti:Sapphire laser and a 25x 0.95 NA immersion lens. ZsGreen was excited with a wavelength of 880 nm.
 \vspace{-.3cm}
\subsection{Dataset}
\label{sec:dataset}
The 3D movies of microglia were imaged over 13 minutes with z-stacks taken at one minute intervals, containing single or multiple microglia per field of view. Some of the images were cropped from a larger field of view containing about 10 different cells and two images were imaged from a zoomed in view of one individual cell. The images ranged from a horizontal pixel width of .01 $\mu$m and a vertical pixel width of .01 $\mu$m to horizontal pixel width of .2 $\mu$m and a vertical pixel width of .2 $\mu$m. In the 3D images, there is variation in the intensity contrast throughout the cell, non-structural noise, and fluorescence bleeding through z-stack due to  the lengthy imaging technique which makes it difficult to visualize and process. The images were preprocessed using histogram equalization, which increased the intensity throughout the cell but further increased noise in the background. 

In our experiments, the segmentation at time \textit{t} = 1 was attained using the coupled tubularity flow field and blob flow field (Tuff-Bff) algorithm \cite{ly2018coupled}.

\subsection{Performance Evaluation}
We use a dataset consisting of 3D images of microglia over a time of 13 minutes, as described in Section \ref{sec:dataset} . We compare our reconstruction of temporal skeleton results with an automatic skeleton reconstruction method called Skel2Graph3D, which requires a 3D segmentation of the original image at each time and was used to reconstruct osteocyte cells \cite{kollmannsberger2017small}. This was done as a comparison, because osteocytes are similar in morphology to microglia. Their 3D skeletonization function is based on a medial axis thinning algorithm \cite{lee1994building, homann2007implementation}, but the Skel2Graph3D algorithm iteratively prunes the skeleton and converts it into graph representation. The ground truth was attained using the Simple Neurite Tracer in ImageJ, which is a semi-automatic tracing software \cite{longair2011simple}. We compare the accuracy of  the \textit{Hieroglyph} results and the state-of-the-art comparisons with the ground truth. We note that even the ground truth may have user error due to background noise and intensity inhomogeneity throughout the object of interest. 

From Figure \ref{fig:results}, \textit{Hieroglyph} has a consistent structure over time. The algorithm maintains consistency by its working principle. But the spatiotemporal localization of the consistent skeletons are obtained within a margin of error.  Our temporal results are based on the result of the prior image but we can see that the skeleton over time changes as the cell changes. We use a hierarchical weighting method to compare the accuracy scores. The branches in each hierarchy are counted and the true and false count is attained by comparing with the ground truth and the accuracy $\frac{TP}{TP+FP+FN}$ is attained for each hierarchy. The final accuracy is found by giving a higher weight to the hierarchies closest to the soma, or the primary branches. The weight is established with the factorial of the maximum number of hierarchies times the accuracy at each hierarchy given by $A_{total}=H_{\max}^{gt}!\sum_{n=0}^i{A}_{Hn}$ where $H_{gt}$ is the number of hierarchies in the ground truth and i is the total number of hierarchies in the test image. This final accuracy for \textit{Hieroglyph} for the first time stack is 55\% for \textit{Hieroglyph}, and 34\% for Skel2Graph. Since the accuracy of the skeleton over time is dependent on the accuracy of the first skeleton, we consider additional measurements for comparison. 

\begin{figure}
	\centering
  \includegraphics[width=\linewidth]{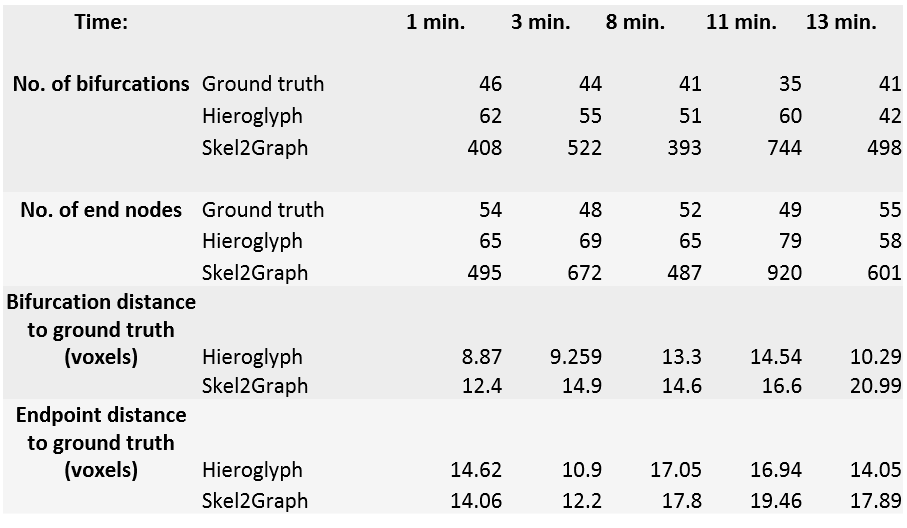}
  \caption{\small{Structural measurements compared to the ground truth.}}
  \label{fig: table}
  	\vspace{-0.5cm}
\end{figure} 

\begin{figure}[h]
	\centering
  \includegraphics[width=.85\linewidth]{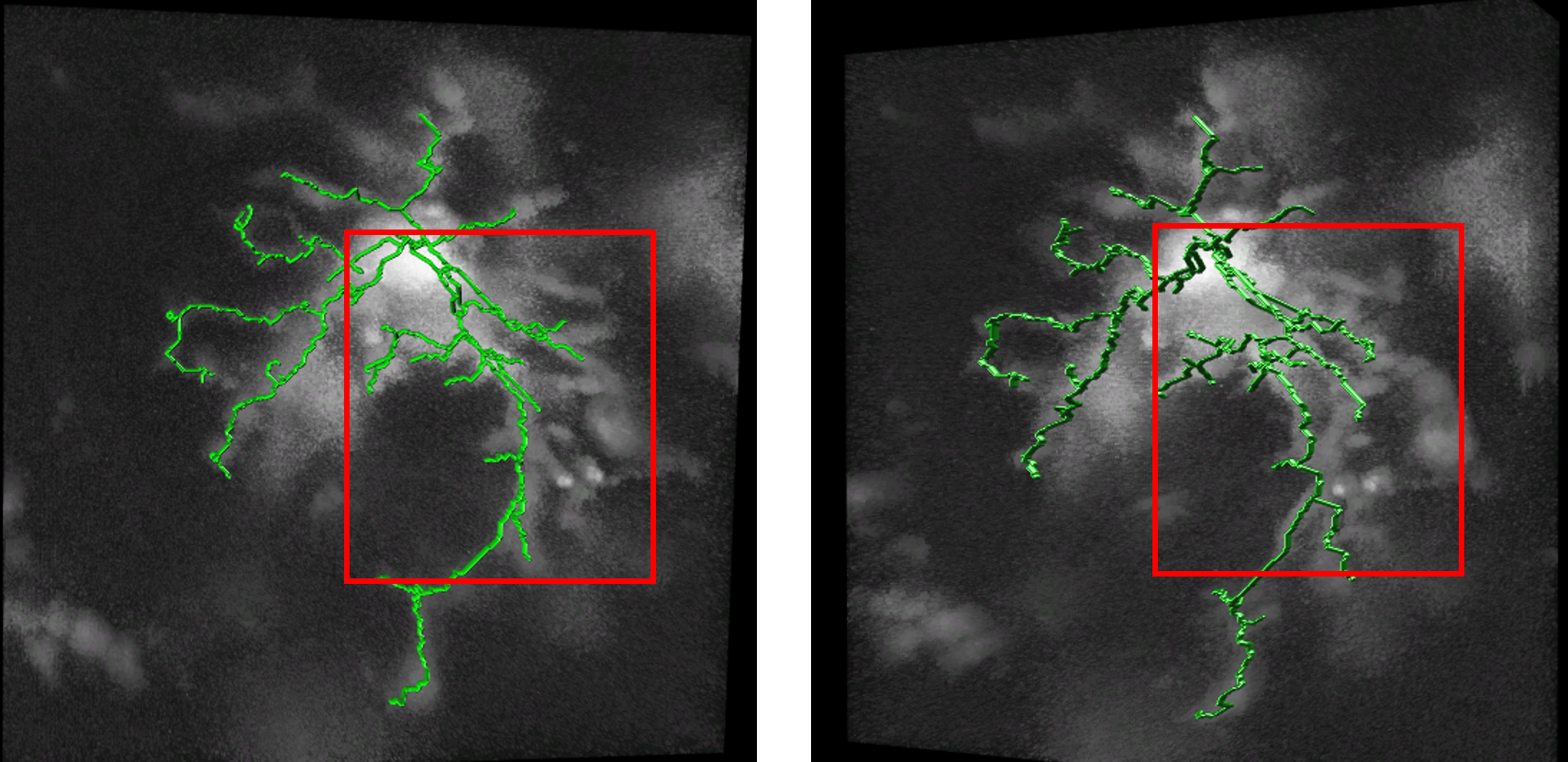}
  \caption{\small{A skeletonization of a glial cell in one orientation (left) and a 3D rotation of the same (right). The rotated view reveal branches not shown in the original view motivating 3D analysis.}}
  \label{fig:rotate}
  	\vspace{-0.5cm}
\end{figure}

We measure the number of bifurcation points and number of terminal nodes as well as the distance between the results and the ground truth, as shown in Figure \ref{fig: table}.  The number of bifurcation points and terminal points in \textit{Hieroglyph} results remain consistent with the number of bifurcation points in the skeletons from the ground truth. The Skel2Graph has a significantly greater number of bifurcation points and endpoints due to the loops.  The distance between the bifurcation points and endpoints of the ground truth's and that of  \textit{Hieroglyph} and Skel2Graph are calculated. The measurements show how structurally similar the resulting skeletons are compared to the ground truth skeleton.  Figure \ref{fig:rotate} shows that a 3D skeleton rotated to make processes extending in the z-direction more visible. Some processes described may not be visible or be accurately distinguished in a 2D image.  This proves the necessity of a 3D skeleton for morphological and motility analysis.

\Section{Conclusion}
In this paper, we proposed an automatic temporal 3D skeletonization  method for glia images. We are able to use the \textit{a priori} information from preceding skeletons to derive subsequent ones. The method is hierarchical since the skeletonization and graph matching are performed in segments starting at the soma and extending to the endpoints of the processes. \textit{Hieroglyph} attained consistent skeleton structures over time.  While our method performed better than the state of the art, it could be further improved to account for the addition and deletion of glia processes over time. \textit{Hieroglyph} does provide rich information for the analysis of glia motility in homeostasis and also in impaired states. 

\clearpage

\bibliographystyle{IEEEtran}
{\small
\bibliography{HGlyph}}
\end{document}